\documentclass{article}[11 pt]
\usepackage{authblk} 
\usepackage{graphicx} 
\usepackage{amsmath,amssymb}
\usepackage{amsthm}
\usepackage{natbib}
\usepackage{verbatim}
\usepackage{multirow}
\usepackage[capposition=top]{floatrow} 
\usepackage{setspace}
\usepackage{silence}
\usepackage{hyperref}
\usepackage{makecell}
\usepackage{float}
\newcommand{\double}{\baselineskip 30pt}
\newcommand{\single}{\baselineskip 15pt}
\newcommand{\bbeta}{\boldsymbol{\beta}}
\newcommand{\boldeta}{\boldsymbol{\eta}}
\newcommand{\pperp}{\perp \!\!\! \perp}
\DeclareMathOperator{\logit}{logit}
\DeclareMathOperator{\minimize}{minimize}

\title{Improving optimal subsampling through stratification}
\author[1,*]{Jasper~B. Yang}
\author[2]{Thomas Lumley}
\author[3]{Bryan~E. Shepherd}
\author[1,4]{Pamela~A. Shaw}

\affil[1]{Department of Biostatistics, University of Washington, Seattle, WA, USA}
\affil[2]{Department of Statistics, University of Auckland, Auckland, New Zealand}
\affil[3]{Department of Biostatistics, Vanderbilt University, Nashville, Tennessee, USA}
\affil[4]{Biostatistics Division, Kaiser Permanente Washington Health Research Institute, Seattle, WA, USA}

\affil[*]{Corresponding Author Email: jbyang@uw.edu}
 \renewcommand\Authands{ and }

\date{\today}
\usepackage[dvipsnames]{xcolor}
\begin{document}
\doublespacing
\maketitle
\newpage

\label{firstpage}

\begin{abstract}
    Recent works have proposed optimal subsampling algorithms to improve computational efficiency in large datasets and to design validation studies in the presence of measurement error. Existing approaches generally fall into two categories: (i) designs that optimize individualized sampling rules, where unit-specific probabilities are assigned and applied independently, and (ii) designs based on stratified sampling with simple random sampling within strata. Focusing on the logistic regression setting, we derive the asymptotic variances of estimators under both approaches and compare them numerically through extensive simulations and an application to data from the Vanderbilt Comprehensive Care Clinic cohort.  Our results reinforce that stratified sampling is not merely an approximation to individualized sampling, showing instead that optimal stratified designs are often more efficient than optimal individualized designs through their elimination of between-stratum contributions to variance. These findings suggest that optimizing over the class of individualized sampling rules overlooks highly efficient sampling designs and highlight the often underappreciated advantages of stratified sampling.
\end{abstract}

\maketitle

\section{Introduction}
\label{s:intro}

Large observational datasets have become increasingly common resources for applied research across many domains, including medicine, public health, economics, and the social sciences. One prominent example is electronic health records (EHRs), which are collected as part of routine clinical care and are relatively inexpensive compared to traditional study cohorts \citep{jensen2012mining, Lee2020}. These data sources typically contain a wide range of information, from patient demographics to medical histories and laboratory results, making them powerful tools for studying large, diverse populations over extended periods of time. However, their use for statistical analysis also presents important challenges. First, the large size of many modern EHR-based datasets can create computational obstacles, making it infeasible to apply maximum likelihood estimation directly to the full cohort \citep{wang2018optimal, keret2023analyzing}. Second, EHRs are designed primarily for clinical and administrative purposes rather than research, so the data often contain substantial measurement error and missingness \citep{botsis2010secondary,Giganti2020,Shepherd2023}. Similar issues arise in many other observational data sources and can introduce bias or reduce efficiency if not appropriately accounted for \citep{Keogh2020}.

A common aim in observational studies using these large datasets is to efficiently estimate a population-level regression parameter. Given the computational and data-quality challenges, a common strategy towards this aim is to select a subsample from the large cohort, collect error-free data for that subsample if necessary, and then compute an estimator for a population parameter using the subsample. Effective implementation of this strategy involves choosing a subsample so that the resulting estimator is as efficient as possible. To this end, \cite{wang2018optimal} developed an algorithm to compute optimal subsampling designs when variables of interest are observed for the entire cohort. In the logistic regression setting with independent, with-replacement sampling, they derived optimal sampling probabilities which can be shown to be equivalent to sampling proportional to each unit’s influence function. \cite{keret2023analyzing} established an analogous result in the Cox regression setting. For logistic regression when the true outcome is not available but only a surrogate (e.g., an error-prone proxy), \cite{marks2025optimal} proposed a two-step design that uses a pilot sample to approximate the optimal probabilities of \cite{wang2018optimal}. In a similar vein, \cite{Wang2023} recently proposed a general framework for optimal subsampling in semi-parametric models. Collectively, these works show that individualized, probability-weighted subsampling schemes can substantially improve efficiency compared to simple random sampling or case-control sampling.

Another line of research focuses on optimal subsampling designs that use stratified sampling. Stratified sampling, long established in survey literature \citep{Neyman1934, Cochran1977}, partitions the cohort into strata, estimates parameters within strata, and then aggregates results into a final cohort-level estimator. This design removes between-stratum variability from the estimator. When stratification is informative, meaning within-stratum variance is low and between-stratum variance is high, stratified estimators can achieve substantial efficiency gains compared to non-stratified approaches. \cite{Chen2020}, \cite{Han2021a} and \cite{rivera2022} showed that under stratified sampling, optimal subsampling probabilities can be obtained by applying Neyman allocation to influence functions.

In this article, we compare these parallel approaches to optimal subsampling, focusing on the logistic regression settings presented by \cite{wang2018optimal} and \cite{marks2025optimal}. In doing so, we highlight the fact that optimal subsampling is more than just selecting optimal probabilities, rather it is about designing an optimal subsampling scheme. The rest of the paper is organized as follows: In Section \ref{s:Methods}, we outline the optimal sampling problem, derive forms for the variances of different proposed approaches, and compare them analytically in specific settings. In Section \ref{s:SimStudy}, we present a simulation study, modelled after those of \cite{wang2018optimal} and \cite{marks2025optimal}, to compare the designs of interest across a variety of data-generating scenarios. In Section \ref{s:DataExample}, we further demonstrate the differences between these approaches using a dataset from the Vanderbilt Comprehensive Care Clinic (VCCC) study, and we discuss the major findings and make recommendations for future studies in Section \ref{s:Discussion}.

\section{Optimal subsampling for logistic regression}
\label{s:Methods}

\subsection{Setup}

Consider the setting where $\mathbf{y} = (y_1, ..., y_N)^T \in \{0,1\}^N$ is a vector of binary outcomes and $\mathbf{X} = (\mathbf{x_1}, ..., \mathbf{x_N})^T$ is a covariate matrix, including a column of 1's for the intercept, such that $\mathbf{x_i}^T \in \mathbb{R}^{p+1}$. In logistic regression, we seek to estimate $\bbeta = (\beta_0, ... ,\beta_p)$ in the model $\text{logit}(Pr(y_i = 1|\mathbf{x}_i)) = \mathbf{x}_i^T \bbeta$. When $(\mathbf{y}, \mathbf{X})$ are fully observed, this estimation is typically performed via maximum likelihood estimation (MLE), which computes an estimate $\hat{\bbeta}_{\text{MLE}}$ as the solution to the score equation
$$
\mathbf{0} = \dot \ell(\hat{\bbeta}_\text{MLE}) = \frac{1}{N}\sum_{i=1}^N (y_i - p_i(\hat{\bbeta}_{\text{MLE}}))\mathbf{x}_i,
$$
where $p_i(\bbeta) = \text{expit}(\mathbf{x}_i^T \boldsymbol{{\beta}})$. Under regularity conditions, $\hat{\bbeta}_\text{MLE}$ is asymptotically linear and admits the form
\begin{equation}\label{eq:ALMLE}
\hat{\bbeta}_\text{MLE} = \bbeta + \frac{1}{N}\sum_{i=1}^N \underbrace{\mathbf{M}_0(\bbeta)^{-1} (y_i - p_i(\bbeta))\mathbf{x}_i}_{h_i(\bbeta, \boldeta)} + o_p(N^{-1/2}),
\end{equation} where $  \boldeta := \mathbf{M}_0(\bbeta) =  \mathbb{E}[\mathbf{M}_x(\bbeta)]$, $\mathbf{M}_x(\bbeta) = \frac{1}{N}\sum_{j=1}^N w_j(\bbeta) \mathbf{x}_j \mathbf{x_j}^T$, and  $w_i(\bbeta) = p_i(\bbeta)(1-p_i(\bbeta))$. Here, $h_i(\bbeta, \boldeta)^T$ is the $i$-th row of the $n \times (p+1)$ matrix of influence functions $\mathbf{H}(\bbeta, \boldeta)$.


In medical research, this standard procedure is often complicated by the fact that $(\mathbf{y}, \mathbf{X})$ are not fully observed. For instance, \cite{marks2025optimal} consider the setting where $\mathbf{y}$ is not observed at all, with a surrogate $\mathbf{s}$ observed in its place. This standard procedure can also be complicated by computational challenges when $N$ is very large, as in the setting considered by \cite{wang2018optimal}. In either case, an effective analysis strategy is to obtain $y$ for a subset of $n < N$ units, where $n$ is fixed by budget or computational restrictions. Without assuming a model on the relationship between $(\mathbf{y}, \mathbf{X})$ and any available surrogates, a standard way to estimate $\bbeta$ is to compute an estimate $\tilde{\bbeta}$ as the solution to the weighted score equation: 
\begin{equation}\label{eq:weightedscore}
\mathbf{0} = \frac{1}{N}\sum_{i=1}^N \frac{R_i}{\pi_i}(y_i - p_i(\tilde{\bbeta}))\mathbf{x}_i,
\end{equation}
where $\pi_i$ is the inclusion probability for unit $i$ in the validation subsample and $R_i$ is the binary indicator for inclusion in the subsample. Under standard regularity conditions and assuming $\pi_i > \delta > 0$, $N \rightarrow \infty$, $n \rightarrow \infty$, and $n/N \rightarrow \gamma \in (0,1)$, then $\tilde{\bbeta}$ is asymptotically linear and can be expressed as a weighted version of Equation \ref{eq:ALMLE}:

\begin{equation}\label{eq:ALsubsample}
\tilde{\bbeta} = \bbeta + \frac{1}{N}\sum_{i=1}^N \frac{R_i}{\pi_i} h_i(\bbeta, \boldeta) + o_p(N^{-1/2}).
\end{equation}

In this work, we consider the task of designing a subsampling scheme that minimizes the trace of the asymptotic variance of $\tilde{\bbeta}$ given the available data, corresponding to an A-optimality criterion \citep{Chan1982}. 
Combining Equations \ref{eq:ALsubsample} and \ref{eq:ALMLE}, we see that
\begin{equation}\label{eq:AsympLinComb}
    \tilde{\bbeta} = \bbeta + \frac{1}{N}\sum_{i=1}^N \left(\frac{R_i}{\pi_i} - 1\right)h_i(\bbeta, \boldeta) + \left(\hat{\bbeta}_\text{MLE} - \bbeta\right) + o_p(N^{-1/2}).
\end{equation}
The only component of the asymptotic variance that is affected by the subsampling design is the first term. Our design objective is thus to minimize the variance contributed by this subsampling term conditional on the available data. Importantly, this task involves more than just
selecting $\pi_i$'s. Equations \ref{eq:ALsubsample} and \ref{eq:AsympLinComb} show that given $\mathbf{H}(\bbeta, \boldeta)$, the variance comes from the sample membership indicators $R_i$. These Bernoulli random variables clearly satisfy
$\mathbb{E}[R_i] = \pi_i,$ and $\text{Var}(R_i) = \pi_i(1-\pi_i)$, but the
asymptotic variance of the sum $\tilde{\bbeta}$ will also depend on
$\text{Cov}(R_i, R_j)$ for $i \neq j$. Negative covariances will
decrease the asymptotic variance of $\tilde{\bbeta}$.

\subsection{Optimal individualized sampling}

In deriving the optimal subsampling scheme, \cite{marks2025optimal} and \cite{wang2018optimal} find the values of $\pi_i$ such that the variance of $\tilde{\bbeta}$ is minimized when $n$ subsamples are selected independently according to the values of $\pi_i$. We call this class of sampling schemes, which require that $R_i \pperp R_j$ for $i \neq j$ in Equation \ref{eq:ALsubsample}, individualized sampling. The class of individualized sampling schemes includes Poisson sampling, which draws each $R_i$ from a $ \text{Bernoulli}(\pi_i)$ distribution and hence does not control the subsample size but does prevent observations from being sampled multiple times. It also includes with-replacement sampling, which controls the sample size at $n$ but allows the same unit to appear multiple times in the $n$ samples. Allowing $\pi_i$'s to vary at the individual level and sampling without replacement while maintaining a fixed sample size is not considered in this work due to the high computational burden and narrow efficiency gains over with-replacement sampling \citep{tille2006sampling}. The optimal individualized sampling algorithms presented by \cite{wang2018optimal} and \cite{marks2025optimal} use with-replacement sampling, but \cite{wang2021comparative} point out that Poisson sampling can be slightly more efficient for the same sampling probabilities if one is willing to allow $n$ to sometimes exceed the budgeted sample size. 

\cite{wang2018optimal} derive the optimal individualized sampling probabilities motivated by the A-optimality criterion (OSMAC) in the setting where $(\mathbf{y},\mathbf{X})$ is observed for all $N$ units as
\begin{equation}\label{eq:OSMAC1}
   \pi_{i,\text{OSMAC}} =  n \frac{|y_i - p_i(\hat{\bbeta}_\text{MLE})| \|\mathbf{M_x}(\hat{\bbeta}_\text{MLE})^{-1} \mathbf{x}_i\|}{\sum_{j=1}^N|y_j - p_j(\hat{\bbeta}_\text{MLE})| \|\mathbf{M_x}(\hat{\bbeta}_\text{MLE})^{-1} \mathbf{x}_j\|}.
\end{equation}
Note that this expression for $\pi_{i, \text{OSMAC}}$ differs from the form in \cite{wang2018optimal} by a factor of $n$, ensuring that $\sum_{i=1}^N\pi_{i, \text{OSMAC}} = n$, which is a common convention in survey sampling. Functionally, the forms are equivalent. Comparing Equation \ref{eq:OSMAC1} to the influence functions in Equation \ref{eq:ALMLE}, we see that the optimal individualized sampling probabilities are proportional to the Euclidean norms of plug-in estimates of the influence functions. Hence, 
\begin{equation}\label{eq:OSMAC}
   \pi_{i,\text{OSMAC}} /n =  \frac{\|h_i(\hat{\bbeta}_\text{MLE}, \hat{\boldeta})\|}{\sum_{j=1}^N \|h_j(\hat{\bbeta}_\text{MLE}, \hat{\boldeta})\|} \propto \|h_i(\hat{\bbeta}_\text{MLE}, \hat{\boldeta})\| \overset{p} {\rightarrow} \|h_i(\bbeta, \boldeta)\|,
\end{equation}
where $\hat{\boldeta} = \mathbf{M}_x(\hat{\bbeta}_\text{MLE})$, and the convergence in probability follows from $\hat{\bbeta}_\text{MLE} \overset{p}{\rightarrow} \bbeta$ and $\hat{\boldeta} \overset{p}{\rightarrow} \boldeta$ since $\|h_i(\cdot) \|$ is continuous.

By Equation \ref{eq:ALsubsample}, $\tilde{\bbeta}$ can be represented asymptotically as a weighted mean of influence functions, so its asymptotic variance under Poisson sampling follows from the standard variance formula for a Poisson-sampled mean. Under $\pi_{i,\text{OSMAC}}$, this yields
\begin{equation}\label{eq:OSMACvar}
\text{Var}_\text{Pois}\left(\tilde{\bbeta}_\text{OSMAC} \big| (\mathbf{y}, \mathbf{X}) \right) = \frac{1}{N^2}\sum_{i=1}^N \left(
\frac{\sum_{j=1}^N \|h_j(\bbeta, \boldeta)\|}{n\|h_i(\bbeta, \boldeta)\|} - 1\right) h_i(\bbeta, \boldeta) h_i(\bbeta, \boldeta)^T.
\end{equation}

For the case where only $(\mathbf{s}, \mathbf{X})$ are observed for all $N$ units, \cite{marks2025optimal} derive an optimal surrogate-assisted sampling strategy (OSSAT), which uses $\mathbf{s}$ and the relationship between $(\mathbf{s}, \mathbf{X})$ and $\mathbf{y}$ learned from a sub-optimally sampled pilot study of size $n_1 < n$ to determine the optimal individual sampling probabilities $\pi_{i,\text{OSSAT}}$ for the remaining $n_2 = n - n_1$ samples. Their probabilities take the form

\begin{equation}\label{eq:OSAT}
   \pi_{i,\text{OSSAT}} =  n_2 \frac{\sqrt{p_i^s -2p_i^sp_i + p_i^2} \|\mathbf{M_x}(\hat{\bbeta}_\text{MLE})^{-1} \mathbf{x_i}\|}{\sum_{j=1}^N\sqrt{p_j^s -2p_j^sp_j + p_j^2} \|\mathbf{M_x}(\hat{\bbeta}_\text{MLE})^{-1} \mathbf{x_j}\|}, 
\end{equation}
where $p_i = Pr(y_i = 1 | \mathbf{x}_i)$ and $p_i^s = Pr(y_i = 1 |s_i, \mathbf{x}_i)$. The aim of this procedure is to approximate $\pi_{i, \text{OSMAC}}$, and indeed the authors remark that they are approximately equal when $p_i^s$ is well-estimated.

\subsection{Optimal stratified sampling}

Individualized sampling schemes comprise only a subset of possible probability subsampling schemes. Another approach is stratified sampling. Under stratified sampling, the population is partitioned into $K$ subpopulations. Then, samples are selected within each stratum $k = 1, ..., K$ according to a design specific to that stratum, and selection in $k$ is independent from selection in another stratum $k' \neq k$. A common approach is to use simple random sampling (SRS) in each stratum. Write $N_k$ for the number of elements in the population in stratum $k$, $n_k$ for the sample size in stratum $k$, and $I_k$ for the set of indices of elements in stratum $k$. Under stratified sampling with SRS in each stratum, we have $\pi_i = \frac{n_k}{N_k}$ for $i \in I_k$. 

Importantly, this approach is different from setting $\pi_i = \frac{n_k}{N_k}$ in an individualized sampling scheme because here $\text{Cov}(R_i, R_j) = \frac{n_k(n_k-1)}{N_k(N_k-1)} - \frac{n_k^2}{N_k^2}< 0$ if $i,j \in I_k$ and 0 otherwise,
so sample membership indicators are negatively correlated within strata and uncorrelated across strata. Again noting the asymptotic representation of $\tilde{\bbeta}$ as a weighted mean and appealing to the standard stratified sampling variance formula, we find that asymptotically
\begin{align}
\text{Var}_\text{Strat}\left(\tilde{\bbeta} \big| (\mathbf{y}, \mathbf{X}) \right) &= \frac{1}{N^2}\left(\sum_{k=1}^K N_k^2 V_{h,k}/n_k - \sum_{k=1}^K N_k V_{h,k}\right) = \frac{1}{N^2}\sum_{k=1}^K N_k^2\frac{1-\frac{n_k}{N_k}}{n_k} V_{h,k} ,\label{eq:STRATVAR1}
\end{align}
where $V_{h,k}$ is the variance matrix of $h(\bbeta)$ within stratum $k$. Notably, the overall variance never appears in Equation \ref{eq:STRATVAR1}. \cite{rivera2022} show that if the full data are known, then the optimal values of $\pi_i$ in this case are obtained by selecting $n_k$'s so that
\begin{equation}\label{eq:NeymanAll} n_k = n\frac{N_k\sqrt{\text{Tr}(V_{h,k})}}{\sum_{k'=1}^KN_{k'}\sqrt{\text{Tr}(V_{h,k'})}}, \end{equation}
 where $\text{Tr}(\cdot)$ is the trace operator. 
 This expression is only valid if $\text{Tr}(V_{h,k}) > 0$. If it is zero, then stratum $k$ contributes no variance to the estimator regardless of its sample size, and it would be reasonable to set  $n_k$ to 1 in such a case. Applying this design to Equation \ref{eq:STRATVAR1} leads to an asymptotic variance of
\begin{equation}\label{eq:STRATVAR}
\text{Var}_\text{Strat}\left(\tilde{\bbeta}_\text{Neyman} \big| (\mathbf{y}, \mathbf{X}) \right) = \frac{1}{N^2}\sum_{k=1}^K N_k V_{h,k}\left(\frac{\sum_{k'=1}^KN_{k'}\sqrt{\text{Tr}(V_{h,k'})}}{n\sqrt{\text{Tr}(V_{h,k})}} - 1\right).
\end{equation}
As with the OSMAC design, this optimal sampling scheme can only be implemented when $(\mathbf{y}, \mathbf{X})$ are observed for all $N$ units. When $(\mathbf{s}, \mathbf{X})$ are observed instead, a practical alternative is to approximate the optimal design by collecting values of $\mathbf{y}$ from a sub-optimally sampled pilot study of size $n_1 < n$ and using them to estimate $V_{h,k}$. A reasonable pilot study design is to conduct stratified sampling using the first equality in Equation \ref{eq:NeymanAll}, only replacing $V_{h,k}$ with the within-stratum variance of the influence functions corresponding to the MLE computed on the whole cohort using $\mathbf{s}$ instead of $\mathbf{y}$ in the surrogate outcome setting, or some other estimate of the influence functions in the computational feasibility setting. The remaining $n - n_1$ samples are then allocated so that the overall sample of size $n$ is allocated with $n_k$'s according to Equation \ref{eq:NeymanAll}, replacing the $V_{h,k}$'s with their pilot-based estimates. This adaptive approach is described in detail in \cite{Mcisaac2015} and \cite{Yang2025a}.

\subsection{Comparing individualized vs. stratified sampling variances}

 At a high level, the difference between the optimal variances of individualized and stratified sampling schemes, presented in Equations \ref{eq:OSMACvar} and \ref{eq:STRATVAR} respectively, involves a trade-off between individualized sampling's ability to uniquely select each $\pi_i$ and stratified sampling's ability to only include within-stratum variation. To illustrate this trade-off, we consider two extremes. First, suppose stratification is perfectly informative, so that $V_{h,k} = 0$ for all $k = 1, \ldots, K$. This occurs, for example, when $\mathbf{X}$ consists of discrete covariates, $(\mathbf{y}, \mathbf{X})$ is observed for all $N$ units, and strata are formed by exact levels of $(X,Y)$. In this case, any allocation with $n_k > 0$ yields $\text{Var}_\text{Strat}\left(\tilde{\bbeta} \big| (\mathbf{y}, \mathbf{X}) \right) = 0$ in Equation \ref{eq:STRATVAR1}. In contrast, the optimal individualized scheme leads to $\text{Var}_\text{Pois}\left(\tilde{\bbeta}_\text{OSMAC} \big| (\mathbf{y}, \mathbf{X}) \right) >0$ in Equation \ref{eq:OSMACvar}. Thus, when stratification is highly informative, stratified sampling is more efficient than individualized sampling. At the opposite extreme, if stratification is completely uninformative, so that each $V_{h,k}$ equals the overall variance $V_h$ in all strata, then optimal individualized sampling achieves lower variance than optimal stratified sampling (see proof in Supplement).

Most practical settings fall between these two extremes. A useful heuristic is that informative stratification can yield substantial gains in efficiency. In particular, when influence functions are known or can be well-approximated, stratifying on their quantiles is a useful way to construct informative strata \citep{Amorim2021}. This is the strategy we adopt in our simulations.



\section{Simulation Study}
\label{s:SimStudy}


\subsection{Simulation setting}
\label{ss:Simset}

We consider simulation settings that are nearly identical to those of \cite{marks2025optimal} and \cite{wang2018optimal}, only we separately consider settings with three and seven covariates. As in those works, we consider generating covariates $\mathbf{X}$ from six different distributions, which are described in detail in \cite{gelman1995bayesian}:

\begin{enumerate}
    \item \emph{zeroMean Normal:} $\mathbf{x} \sim N(0, \Sigma)$, where $\Sigma_{ij} = 0.5^{I(i\neq j)}$ for the indicator function $I()$.
    \item \emph{rareEvent:} $\mathbf{x} \sim N(-1.6, \Sigma)$. 
    \item \emph{unequalVar:} $\mathbf{x} \sim N(0, \Sigma^*)$ where $\Sigma^*_{ij} = 0.5$ for $i \neq j$ and $\Sigma^*_{ii} = 1/i^2$.
    \item \emph{mixNormal:} $\mathbf{x}$ is the mixture of two multivariate normal distributions $0.5N(1, \Sigma)$ and $0.5N(-1, \Sigma)$.
    \item \emph{T3:}  $\mathbf{x} \sim t_3(0, \Sigma)$/10, where $t_3$ is a multivariate $T$ distribution with 3 degrees of freedom.
    \item \emph{Exp:} Each covariate in $\mathbf{x}$ follows an exponential distribution with a rate parameter of 2. 
\end{enumerate}
To further demonstrate the utility of stratified sampling, we also consider a 7th setting with three covariates:

\begin{enumerate}
\setcounter{enumi}{6}
    \item \emph{DiscreteX:} Each covariate in $\mathbf{X}$ is binary.
\end{enumerate}
In each covariate setting, we consider binary outcomes $y_i$ which take value 1 with probability $p_i = \text{expit}(\mathbf{x_i^T \bbeta})$ where for $p \in \{3,7\}$, $\bbeta = (\beta_0, ..., \beta_p) = (0.5,...,0.5)$, except in the \emph{Exp} setting where $\beta_0 = -0.5$. As in \cite{marks2025optimal}, we also consider surrogate outcome variables $s_{i,\text{low}}$ and $s_{i,\text{high}}$, where ``low" and ``high" refer to degrees of differential misclassification compared to the true outcome $y_i$. Specifically, $s_{i,\text{low}}$ is generated with specificity $0.1 I(X_{i,1} < c_1) + 0.8$ and sensitivity $0.04I(X_{i,1} < c_1) + 0.95$, and $s_{i,\text{high}}$ is generated with specificity $0.1I(X_{i,1} < c_1) + 0.6$ and sensitivity $0.05I(X_{i,1} < c_1) + 0.9$. Here, $c_1$ is the 30\% quantile of $X_1$, except in the \emph{DiscreteX} setting where $c_1 = 0.5$.

For each combination of covariate setting, subsample budget $n \in \{800, 1200, 1600\}$, and pilot sample size $n_1 \in \{200, 600\}$, we performed 1,000 simulation iterations. At each iteration, we first generated covariates, outcome, and surrogate outcomes for $N=10,000$ samples. We then conducted subsampling according to each of the following strategies: 
\begin{itemize}
    \item Strategy 1: Case-control sampling using the true outcome $y$.
    \item Strategy 2: Case-control sampling using the surrogate outcome $s$.
    \item Strategy 3: Optimal individualized sampling using the OSMAC probabilities of \cite{wang2018optimal}, which assumes the true outcome $y$ is observed for all $N$ units.
    \item Strategy 4: Optimal individualized sampling with a pilot study of size $n_1$ using the two-step algorithm of \cite{marks2025optimal}, which assumes that $y$ can only be observed through subsampling.
    \item Strategy 5: Optimal stratified sampling using the optimal allocation in Equation \ref{eq:NeymanAll}, estimating the within-stratum variance of influence functions $V_{h,k}$ with the variance of $\mathbf{H}(\hat{\bbeta}_\text{MLE}, \hat{\boldeta})$, which assumes the true outcome $y$ is observed for all $N$ units. This is the stratified sampling analog of Strategy 3. Strata were formed based on values of $\mathbf{y}$ and the 0.2 and 0.8 quantiles of influence functions for $\beta_1$ through $\beta_3$, even for the seven covariate scenario, yielding up to $2\cdot3^3 = 54$ strata.
    \item Strategy 6: Optimal Stratified sampling with a pilot sample of size $n_1$ to estimate $V_{h,k}$ assuming $y$ can only be observed through subsampling, as described by \cite{Mcisaac2015} and \cite{Yang2025a}. Strata were formed as in Strategy 5, only using $\mathbf{s}$ instead of $\mathbf{y}$.
\end{itemize}
More detailed descriptions of each strategy are provided in the Supplemental Materials. Notably, Strategies 1, 3, and 5 are not feasible in the setting where $y$ is unknown, but we included them because they represent the ``best possible" versions of Strategies 2, 4, and 6 respectively. We assessed the performance of each strategy by computing the empirical mean-squared error of the inverse-probability weighted MLE in each setting: $MSE = \frac{1}{1000}\sum_{j=1}^{1000}\|\tilde{\bbeta}_j - \bbeta\|^2$, where $\tilde{\bbeta}_j$ is the estimate for the true $\bbeta$ in the $j$-th simulation iteration. For the \emph{DiscreteX} setting, we also performed an additional simulation where we hold the values of $\mathbf{X}$ constant over all simulation iterations to isolate the variance due to subsampling.

\subsection{Simulation results}\label{ss:SimResults}

\begin{figure}
\centerline{\includegraphics[width=5.8in]{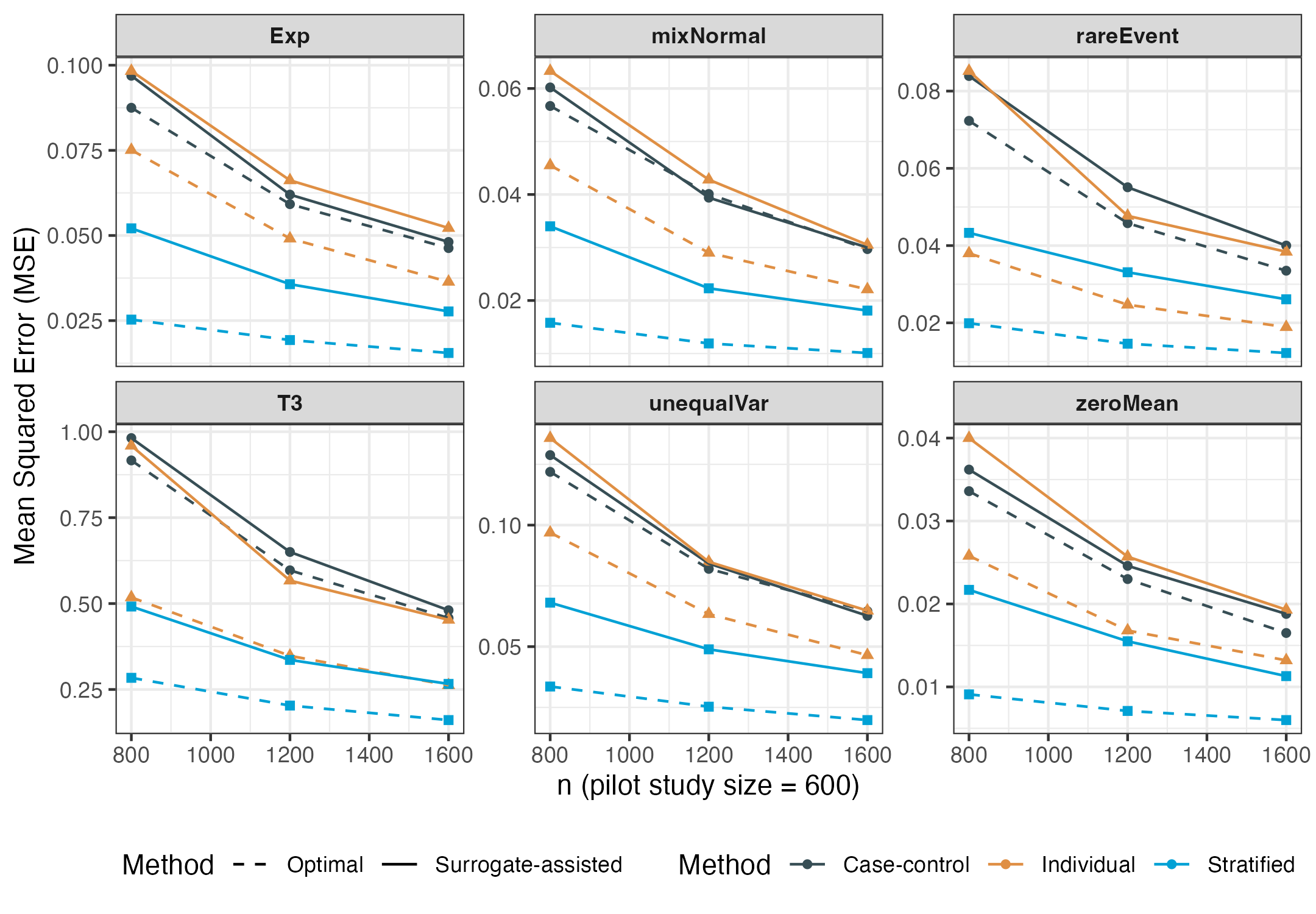}}
\caption{Empirical MSEs of each strategy under data-generating scenarios 1-6 with three covariates and a low level of misclassification for the surrogate. Dashed lines represent strategies that rely on $y$ being known, and solid lines represent strategies that use a surrogate and/or a pilot study. }
\label{fig:1}
\end{figure}

The empirical MSEs for each strategy under data-generating scenarios 1-6 with a pilot sample size of $n_1 = 600$ are shown in Figure \ref{fig:1} for the three covariate case setting. Optimal stratified sampling assuming $y$ is known, which uses Neyman allocation using influence functions from the full-data MLE, achieved the lowest MSE across all six scenarios. The optimal individualized probabilities of \cite{wang2018optimal}, which also assume $y$ is known, consistently outperformed case-control sampling but were always worse than optimal stratified sampling. The individualized and stratified approaches were most similar in the \emph{rareEvent} scenario.

Among the strategies that are feasible when $y$ is unknown (i.e. in the error-prone EHR setting), stratified sampling with a pilot study outperformed both case-control sampling using the surrogate outcome and the optimal individualized sampling approach of \cite{marks2025optimal} in every three-covariate scenario with a pilot size of $n_1 = 600$. With $n_1 = 200$, the stratified sampling approach outperformed the others in the low error setting, but led to the highest MSE in the high error setting (Supplemental Figures S1-S2). This poor performance under a small pilot study aligns with findings and discussions from \cite{Chen2020}. Notably, the stratified sampling with pilot study approach even outperformed the optimal individualized approach assuming a known $y$ in four out of the six settings shown in Figure \ref{fig:1}.

\begin{figure}[!htbp]
\centerline{\includegraphics[width=5.8in]{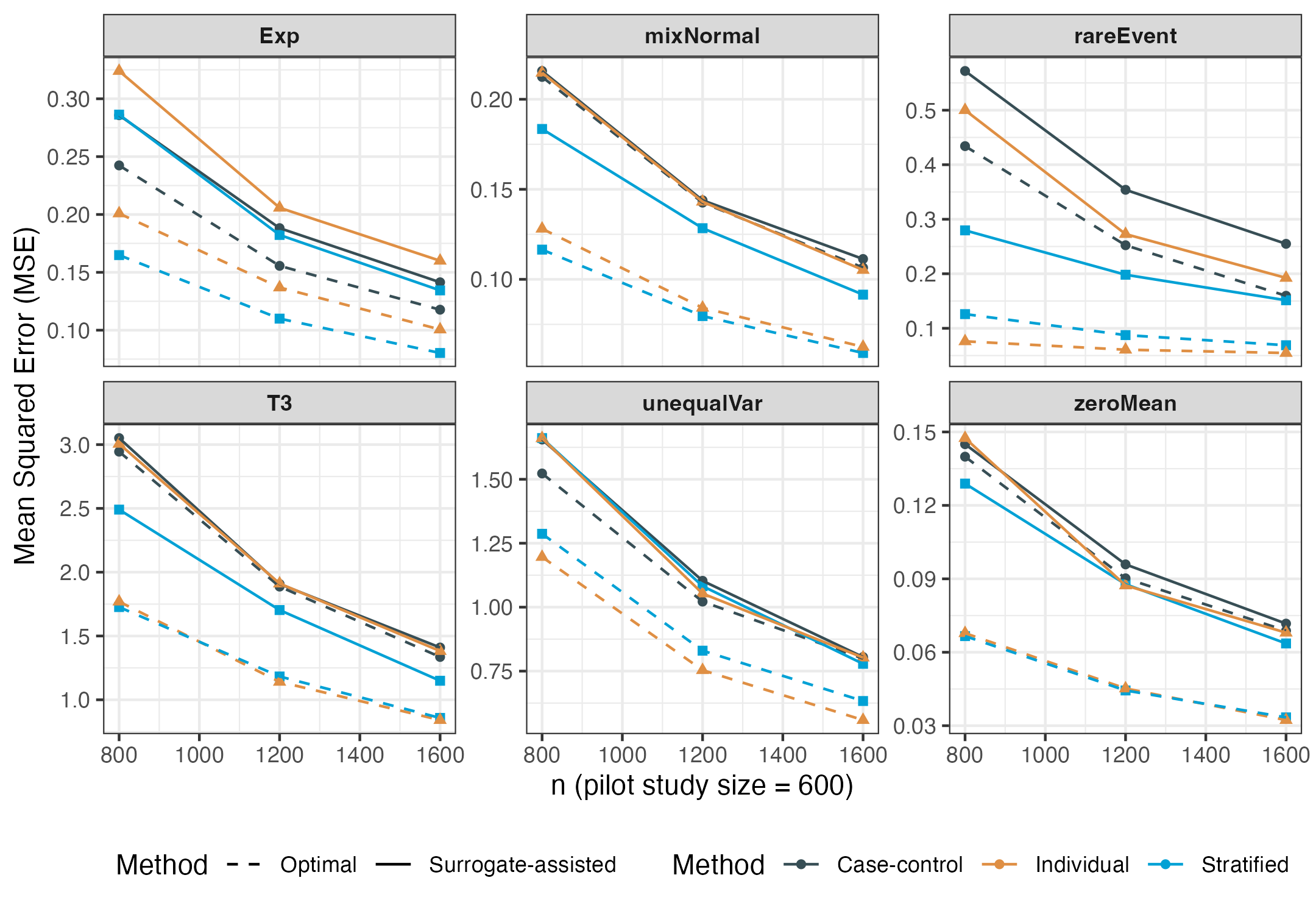}}
\caption{Empirical MSEs of each strategy under data-generating scenarios 1-6 with seven covariates and a low level of misclassification for the surrogate. Dashed lines represent strategies that rely on $y$ being known, and solid lines represent strategies that use a surrogate and/or a pilot study. }
\label{fig:2}
\end{figure}

Figure \ref{fig:2} shows the empirical MSEs in the seven-covariate setting. Here, the advantages of stratified over individualized sampling were more modest because strata were constructed using influence functions for only three of the eight $\beta$’s of interest. In this case, optimal stratified sampling with known $y$ clearly outperformed optimal individualized sampling in only two scenarios, performed equally in two, and was worse in the remaining two. In the measurement error setting, stratified sampling with a pilot study remained the best or nearly the best approach, although the differences among methods were smaller than in the three-covariate setting.

\begin{figure}[!ht]
\centerline{\includegraphics[width=6in]{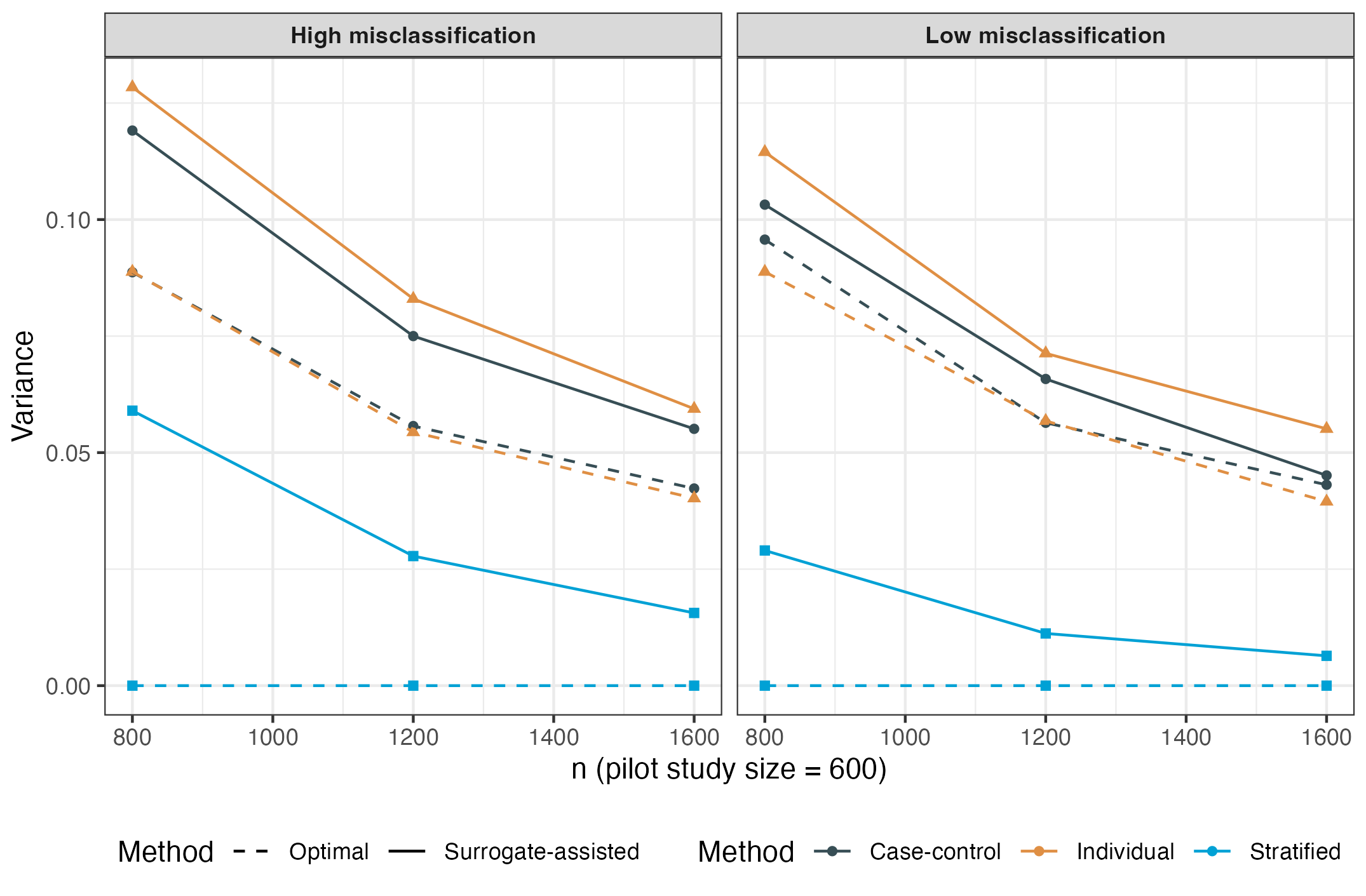}}
\caption{Sum of empirical variances for $\hat{\beta}_0, \hat{\beta}_1, \hat{\beta}_2,$ and $\hat{\beta}_3$ of each strategy under categorical X data-generating scenario and a low level of misclassification for the surrogate. The full data are held constant across each simulation iteration, so the only source of variability is due to subsampling. Dashed lines represent strategies that rely on $y$ being known, and solid lines represent strategies that use a surrogate and/or a pilot study. }
\label{fig:3}
\end{figure}

Figure \ref{fig:3} shows the sum of empirical variances for $\hat{\beta}_1, \hat{\beta}_2, \hat{\beta}_3,$ and  $\hat{\beta}_4$ in the \emph{DiscreteX} data-generating scenario with $X$ fixed across simulation iterations. In this setting, the only source of variability is the selection of the $n$ subsamples. Notably, the optimal stratified sampling strategy in the setting where $\mathbf{y}$ is known yields no variance, meaning it recovers the exact MLE that would have been obtained using the full data despite only using the subsample. The pilot-based stratified approach, designed to approximate this oracle strategy, achieves variances close to zero and outperforms all other feasible methods by a wide margin. In contrast, the optimal individualized approach leads to non-zero variance even when $\mathbf{y}$ is known. This scenario illustrates the efficiency gains attainable with stratified sampling.

\pagebreak

\section{Example: Vanderbilt Comprehensive Care Clinic Study}
\label{s:DataExample}

We also assessed the performance of each optimal sampling strategy using an EHR-derived dataset from the Vanderbilt Comprehensive Care Clinic (VCCC). This dataset contained information on 1,595 people living with HIV who received care at the clinic between 1997 and 2013, and it includes variables collected at the time of antiretroviral therapy (ART) initiation, including age and CD4 cell count, and an indicator for AIDS-defining event (ADE) within 10 years of ART initiation. It was especially suitable for comparing the proposed methods because researchers validated the entire dataset through chart-review, so the dataset contains both validated and error-prone versions of the outcome of interest, ADE within 10 years of ART initiation. Here, ADE within 10 years is a rare event, with a prevalence of 6\%. The error-prone version of this variable, which we use as the surrogate $\mathbf{s}$, has a prevalence of 13\%, and its sensitivity and specificity are 0.83 and 0.90 respectively. Moreover, the sensitivity is 0.72 among people with initial CD4 counts below the 0.3 quantile and 0.9 among people with initial CD4 counts above the 0.3 quantile, suggesting differential misclassification.
Further information on the data and VCCC cohort can be found in \cite{Giganti2020}, \cite{Oh2021a}, and \cite{yang2025optimal}.

We compared the presented methods over 1,000 simulation iterations. At each iteration, we selected $n=200$ subsamples according to each strategy and computed final estimates by solving the weighted score equation in \ref{eq:weightedscore} under the logistic model:
\[
\text{logit}(Pr(\text{ADE10})|\text{Age}, \text{CD4}) = \beta_0 + \beta_\text{Age}\text{Age} + \beta_\text{CD4} \text{CD4}, 
\]
where $\text{ADE10}$ is a binary indicator for ADE within 10 years of ART initiation, Age is age in years at ART initiation, and CD4 is CD4 cells per cubic millimeter at ART initiation. The stratified sampling strategies used 18 strata formed by combinations of ADE status and Age and CD4 partitioned at their 0.2 and 0.8 quantiles. For the strategies that use a pilot study, we varied the pilot study size from $n_1 = 50$ to $n_1 = 125$, conducting 1,000 simulation iterations for each size. 

\begin{figure}[!htpb]
\centerline{\includegraphics[width=6in]{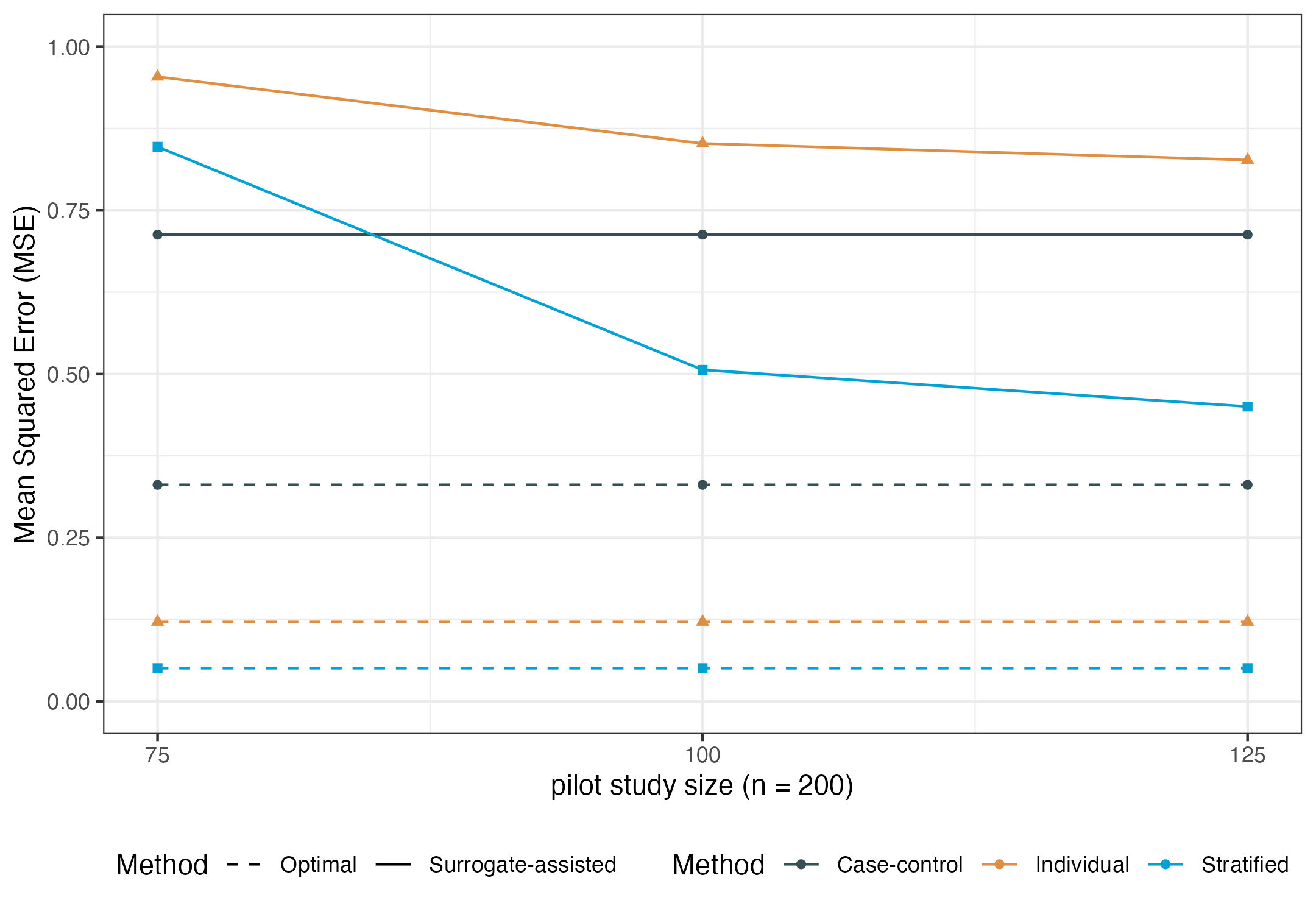}}
\caption{Sum of empirical MSEs for $\hat{\beta}_0, \hat{\beta}_\text{CD4}, \hat{\beta}_\text{Age}$ for each strategy under pilot study sizes of 75, 100, and 125 with the entire subsample size fixed at $n = 200$. Dashed lines represent strategies that rely on $y$ being known, and solid lines represent strategies that use a surrogate and/or a pilot study. }
\label{fig:4}
\end{figure}

The sum of the empirical MSEs of $\beta_0, \beta_\text{Age}$, $\beta_\text{CD4}$ under each strategy at different pilot sample sizes are presented in Figure \ref{fig:4}. To compute the MSEs, we used the MLE estimates that would have been obtained using the entire error-free dataset as the ``truth" for each coefficient, isolating the variance due to subsampling. Among the optimal subsampling strategies that were allowed access to the error-free variables, the stratified approach was slightly more efficient than the optimal individualized sampling approach, and both optimal approaches outperformed case-control sampling using the true outcome. Among strategies that only use the surrogate outcome and/or a pilot study, the stratified approach was the most efficient when the pilot study size was 100 or 125, but when conducted with a pilot study size of 75, it was less efficient than case-control sampling on the surrogate outcome. This result is consistent with recommendations made by \cite{Mcisaac2015} and \cite{Han2021a} for a pilot sample size of size $n_1 = n/2$. This data example provide further evidence for stratified sampling in the case where strata are informative and the parameters for Neyman allocation can be well-approximated.

\section{Discussion}
\label{s:Discussion}

Through analytical derivations, extensive simulations, and a data application, we have shown that optimal stratified subsampling designs can yield more efficient estimators than optimal individualized subsampling designs in logistic regression. In the oracle setting where the full-data influence functions are known, stratified sampling almost always outperforms individualized sampling when the dimension of the influence functions is small enough to permit stratification on their quantiles. The key advantage is that stratification eliminates all between-stratum variance in the final estimator, and this gain typically outweighs the modest loss in efficiency incurred by assigning equal sampling probabilities within strata. In the more realistic setting where the full-data influence functions are unknown and only estimates can be obtained through surrogate and/or a pilot study, we showed that in most cases stratified sampling still performed better. This result is notable because much of the recent literature in optimal subsampling has focused on optimizing over the class of individualized sampling rules.

Despite these benefits, our empirical results and simulations also highlight limitations of stratified sampling. When the parameter of interest is high-dimensional, stratifying on combinations of all influence functions becomes difficult. One workaround, as shown in the seven covariate simulation setting, is to only use a subset of influence functions in the stratification, but this reduces the information captured by the stratification and accordingly decreases efficiency. Further, when the stratification variables are weakly correlated with the true variables, as in the high misclassification simulation setting, the informativeness of the stratification also decreases. A further limitation of the stratified sampling approaches considered in this work is that they use simple random sampling within strata. It is reasonable to expect that some combination of individualized sampling and stratified sampling could further improve these simple stratified sampling designs, but we leave this exploration to future work. Together, these limitations show that optimal individualized sampling algorithms remain valuable in certain settings. 

The results of Section \ref{s:Methods} are based on influence functions and can hence be applied more broadly to any asymptotically linear estimator. Thus, although we focused on the inverse probability weighted logistic regression settings of \cite{wang2018optimal} and \cite{marks2025optimal}, the ideas developed here extend naturally to other contexts, including the survival analysis setting of \cite{keret2023analyzing}, generalized raking setting of \cite{Chen2022}, and more general semi-parametric estimation setting of \cite{Wang2023}. We therefore suggest that stratified sampling be considered in any optimal subsampling algorithm.

\section*{Acknowledgments} This project was supported by the U.S. National Institutes of Health (NIH) grants R37-AI131771 and  P30-AI110527. 
\section*{Data Availability} 
Code to reproduce the simulation results is available on Github at \url{https://github.com/yangjasp/StratifiedSampling}. Data from the example in this paper may be obtained by contacting the corresponding author and appropriate data use agreements. Data are not publicly available due to privacy restrictions.

\def\maintex{}

\bibliographystyle{chicago}
\bibliography{references}
\newpage
\ifdefined\maintex
\renewcommand{\thesection}{S\arabic{section}}
\renewcommand{\thetable}{S\arabic{table}}  
\renewcommand{\thefigure}{S\arabic{figure}}
\renewcommand{\figurename}{Supplemental Figure} 
\renewcommand{\tablename}{Supplemental Table} 
\renewcommand{\theequation}{S\arabic{equation}}
\setcounter{section}{1}
\setcounter{equation}{0}
\setcounter{table}{0}
\setcounter{figure}{0}
\setcounter{tocdepth}{2}
\else
\documentclass{article}
\usepackage{authblk}
\usepackage{multirow}
\usepackage{graphicx} 
\usepackage{subcaption} 

\usepackage{amsmath,amssymb}
\usepackage{amsthm}
\usepackage{natbib}
\usepackage{longtable}
\usepackage{makecell}
\usepackage{array}
\usepackage{silence}
\usepackage{comment}
\usepackage[margin=0.75in]{geometry}
\usepackage[capposition=top]{floatrow} 
\usepackage[colorlinks=true,linkcolor=cyan,urlcolor=cyan]{hyperref} 
\hypersetup{
    colorlinks,
    citecolor=black,
    filecolor=black,
    linkcolor=black,
    urlcolor=black
}
\usepackage{silence} 
\usepackage[dvipsnames]{xcolor}

\usepackage{tikz}
\usepackage{float}
\usetikzlibrary{shapes.geometric, arrows.meta, decorations,decorations.markings}
\tikzstyle{standard} = [rectangle, rounded corners, minimum width=2cm, minimum height=1cm,text centered, draw=black]
\tikzstyle{arrow} = [thick,-latex]

\WarningFilter*{latex}{Text page \thepage\space contains only floats}

\WarningFilter*{latex}{Float too large for page by}

\newcommand{\single}{\baselineskip 15pt}
\DeclareMathOperator{\logit}{logit}
\newcommand{\double}{\baselineskip 30pt}
\newcommand{\bbeta}{\boldsymbol{\beta}}
\newcommand{\boldeta}{\boldsymbol{\eta}}
\newcommand{\pperp}{\perp \!\!\! \perp}
\DeclareMathOperator{\minimize}{minimize}

\renewcommand{\thesection}{S\arabic{section}}
\renewcommand{\thetable}{S\arabic{table}}  
\renewcommand{\thefigure}{S\arabic{figure}}
\renewcommand{\figurename}{Supplemental Figure} 
\renewcommand{\tablename}{Supplemental Table} 
\renewcommand{\theequation}{S\arabic{equation}}
\setcounter{table}{0}
\setcounter{figure}{0}
\setcounter{tocdepth}{2}
\newtheorem{assumption}{Assumption}

\title{Supplementary Materials for ``On the efficiency of stratified approaches to optimal subsampling''}

\title{On the efficiency of stratified approaches to optimal subsampling}
\author[1,*]{Jasper~B. Yang}
\author[2]{Thomas Lumley}
\author[3]{Bryan~E. Shepherd}
\author[1,4]{Pamela~A. Shaw}

\affil[1]{Department of Biostatistics, University of Washington, Seattle, WA, USA}
\affil[2]{Department of Statistics, University of Auckland, Auckland, New Zealand}
\affil[3]{Department of Biostatistics, Vanderbilt University, Nashville, Tennessee, USA}
\affil[4]{Biostatistics Division, Kaiser Permanente Washington Health Research Institute, Seattle, WA, USA}
 \renewcommand\Authands{ and }


\date{\today}

\begin{document}

\maketitle

\tableofcontents
\newpage

\fi

\section*{Supplemental Materials}

\subsection{Proofs}

\subsubsection{Proof of asymptotic variances in Equation \ref{eq:STRATVAR1} and Equation \ref{eq:OSMACvar}}

Consider the asymptoptically linear expansion of $\tilde{\bbeta}$ in Equation \ref{eq:AsympLinComb}:
\begin{equation}\label{eq:SUPAsympLinComb}
\tilde{\bbeta} = \bbeta + \frac{1}{N}\sum_{i=1}^N \left(\frac{R_i}{\pi_i} - 1\right)h_i(\bbeta, \boldeta) + \left(\hat{\bbeta}_\text{MLE} - \bbeta\right) + o_p(N^{-1/2}).
\end{equation}
Conditional on $(\mathbf{y}, \mathbf{X})$, $h_i(\bbeta, \boldeta)$ and $\hat{\bbeta}_\text{MLE}$ are fixed, $R_i$ is random, and $\pi_i = n_k/N_k$ is fixed through our selection of $n_k$. Further, by conditioning on the $N$ values in $(\mathbf{y}, \mathbf{X})$, our subsampling problem is one of sampling from a finite population.

First, consider the Poisson sampling variance in Equation \ref{eq:OSMACvar}. Noting that $\hat{\bbeta}_\text{MLE} - \bbeta$ is fixed conditional on $(\mathbf{y}, \mathbf{X})$ and ignoring the variance contributed by the $o_p(N^{-1/2})$ term, which will not appear in the limiting distribution as $N \rightarrow \infty$, we are left with the variance of $\tilde{\beta}$ being represented by the variance of a sample mean under Poisson sampling from a finite population. Applying well-known variance results from survey sampling gives Equation \ref{eq:OSMACvar} \citep{Cochran1977}.

Next, consider the stratified sampling variance in Equation \ref{eq:STRATVAR1}. Here we only consider the non-trivial case where $V_{h,k} > 0$, since as discussed in the main text $V_{h,k} = 0$ yields an asymptotically uninteresting estimator with no variance. Writing $I_k$ for the set of indices in stratum $k$, we can write Equation \ref{eq:SUPAsympLinComb} as
\begin{align*}
\tilde{\bbeta} &= \bbeta + \frac{1}{N}\sum_{k=1}^K \sum_{i \in I_k} \left(\frac{R_i}{\pi_i} - 1\right)h_i(\bbeta, \boldeta)  + \left(\hat{\bbeta}_\text{MLE} - \bbeta\right) + o_p(N^{-1/2}).
\end{align*}
Again noting that $\hat{\bbeta}_\text{MLE} - \bbeta$ is fixed conditional on $(\mathbf{y}, \mathbf{X})$ and ignoring the variance contributed by the $o_p(N^{-1/2})$ term, we are left with the variance of $\tilde{\beta}$ being represented by the variance of a stratified sample mean from a finite population. Applying well-known results on the variance of a stratified sample mean yields Equation \ref{eq:STRATVAR1} \citep{Cochran1977}.

To see the limiting distribution corresponding to this asymptotic variance and more formally handle the vanishing $o_p(N^{-1/2})$ remainder term, note that
\begin{align*}
\tilde{\bbeta}
&= \bbeta + \sum_{k=1}^K \frac{N_k}{N} \left(\frac{1}{N_k}\sum_{i \in I_k}\frac{R_i}{\pi_i}h_i(\bbeta, \boldeta) - \frac{1}{N_k} \sum_{i  \in I_k} h_i(\bbeta, \boldeta)\right)  + \left(\hat{\bbeta}_\text{MLE} - \bbeta\right) + o_p(N^{-1/2}) \\
&= \bbeta + \sum_{k=1}^K \frac{N_k}{N} \left(\frac{1}{n_k}\sum_{i \in I_k}R_i h_i(\bbeta, \boldeta) - \frac{1}{N_k} \sum_{i  \in I_k} h_i(\bbeta, \boldeta)\right)  + \left(\hat{\bbeta}_\text{MLE} - \bbeta\right) + o_p(N^{-1/2}).
\end{align*}


This form sets up a direct application of the results from situation (b) of \cite{bickel1984asymptotic}, as also discussed in \cite{chen2007asymptotic}, to the second term in the sum. Under their assumptions, extended appropriately to the multivariate case, we have, after an application of Slutsky's theorem to to handle the remainder term from the estimating equations setup that conditional on $(\mathbf{y}, \mathbf{X})$,

\[
\left\{\sum_{k=1}^K \frac{N_k}{N} \left(\frac{1}{n_k}\sum_{i \in I_k}R_i h_i(\bbeta, \boldeta) - \frac{1}{N_k} \sum_{i  \in I_k} h_i(\bbeta, \boldeta)\right)\right\}^T\Sigma^{-1}_{N, \text{Strat}}\sum_{k=1}^K \frac{N_k}{N} \left(\frac{1}{n_k}\sum_{i \in I_k}R_i h_i(\bbeta, \boldeta) - \frac{1}{N_k} \sum_{i  \in I_k} h_i(\bbeta, \boldeta)\right) \overset{d}{\rightarrow} \chi^2_p
\]
where $\Sigma_{N,\text{Strat}}$ is exactly the form of $\text{Var}_{n, \text{Strat}}\left(\tilde{\bbeta} \big| (\mathbf{y}, \mathbf{X}) \right)$ provided in Equation \ref{eq:STRATVAR1}.

For the asymptotic distribution in Equation \ref{eq:OSMACvar}, we refer to the results of \cite{hajek1960limiting}, which, when combined with Slutsky's theorem to handle the remainder from the estimating equations setup and appropriate extensions to the multivariate setting, show that

\[
\left\{\frac{1}{N}\sum_{i=1}^N \left(\frac{R_i}{\pi_i} - 1\right)h_i(\bbeta, \boldeta)\right\}^T\Sigma^{-1}_{N, \text{OSMAC}}\frac{1}{N}\sum_{i=1}^N \left(\frac{R_i}{\pi_i} - 1\right)h_i(\bbeta, \boldeta) \overset{d}{\rightarrow} \chi^2_p
\]
where $\Sigma_{N,\text{OSMAC}}$ is exactly the form of $\text{Var}_{n, \text{OSMAC}}\left(\tilde{\bbeta} \big| (\mathbf{y}, \mathbf{X}) \right)$ provided in Equation \ref{eq:OSMACvar}.



Finally, we note that although the details of the multivariate extensions of these asymptotic distribution results are not provided here, the multivariate versions are not actually required for our main results, as we focus only on the trace of the asymptotic variance. This can be represented element-wise using the univariate forms of these variance formulas.

\subsubsection{Proof that individualized sampling is more efficient than stratified sampling when stratification is useless}

If stratification is useless, then  $\text{Var}_k(h(\bbeta, \boldeta)) = \text{Var}(h(\bbeta, \boldeta))$ $\forall$ $k = 1, ..., K$. Recall that we are conditioning on $N$ observations in the full data, so we are working with finite-population variance. Hence, $V_{h;k} = V_h = \text{Var}(h(\bbeta, \boldeta)) = \frac{1}{N}\sum_{i =1}^N h_i(\bbeta, \boldeta)h_i(\bbeta, \boldeta)^T - \allowbreak \left(\frac{1}{N}\sum_{i = 1}^Nh_i(\bbeta, \boldeta)\right)\left(\frac{1}{N}\sum_{i = 1}^Nh_i(\bbeta, \boldeta)\right)^T = \frac{1}{N}\sum_{i =1}^N h_i(\bbeta, \boldeta) h_i(\bbeta, \boldeta)^T - o_p(N^{-1/2})$.


Taking the traces of Equations \ref{eq:STRATVAR} and \ref{eq:OSMACvar}, we have asymptotically
\begin{align*}
\text{Tr}\left(\text{Var}_\text{Strat}\left(\tilde{\bbeta}_\text{Neyman} \big| (\mathbf{y}, \mathbf{X}) \right) \right) &= \frac{1}{N^2}\sum_{k=1}^K N_k \text{Tr}(V_{h,k})\left(\frac{\sum_{k'=1}^KN_{k'}\sqrt{\text{Tr}(V_{h,k'})}}{n\sqrt{\text{Tr}(V_{h,k})}} - 1\right) \\
&= \frac{1}{N^2}\sum_{k=1}^K N_k \text{Tr}(V_{h})\left(\frac{\sum_{k'=1}^KN_{k'}\sqrt{\text{Tr}(V_{h})}}{n\sqrt{\text{Tr}(V_{h})}} - 1\right) \\
&= \frac{N/n-1}{N^2}\sum_{k=1}^K N_k \text{Tr}(V_{h}) \\
&= \frac{N(N/n-1)}{N^2} \text{Tr}(V_{h}) \\
&= \frac{N(N/n-1)}{N^2} \left(\frac{1}{N}\sum_{i =1}^N \text{Tr}\left(h_i(\bbeta, \boldeta) h_i(\bbeta, \boldeta)^T\right) + o_p(N^{-1/2})\right) \\
&= \frac{N(N/n-1)}{N^2} \left(\frac{1}{N}\sum_{i =1}^N \|h_i(\bbeta, \boldeta)\|^2\right) + o_p(1/N) 
\end{align*}
and 
\begin{align*}
\text{Tr}\left(\text{Var}_\text{Pois}\left(\tilde{\bbeta}_\text{OSMAC} \big| (\mathbf{y}, \mathbf{X}) \right)\right) &= 
\frac{1}{N^2}\sum_{i=1}^N \left(
\frac{\sum_{j=1}^N \|h_j(\bbeta, \boldeta)\|}{n\|h_i(\bbeta, \boldeta)\|} - 1\right) \text{Tr}\left( h_i(\bbeta, \boldeta) h_i(\bbeta, \boldeta)^T \right) \\ 
&= \frac{1}{N^2}\sum_{i=1}^N \left(
\frac{\sum_{j=1}^N \|h_j(\bbeta, \boldeta)\|}{n \|h_i(\bbeta, \boldeta)\|} - 1\right) \| h_i(\bbeta, \boldeta)\|^2 
\end{align*}
This gives asymptotically,
\begin{align*}
    \text{Var}_\text{Pois}\left(\tilde{\bbeta}_\text{OSMAC} \big| (\mathbf{y}, \mathbf{X}) \right) - \\ \text{Var}_\text{Strat}\left(\tilde{\bbeta}_\text{Neyman} \big| (\mathbf{y}, \mathbf{X}) \right) &= \frac{1}{N^2}\sum_{i=1}^N \left(
\frac{\sum_{j=1}^N \|h_j(\bbeta, \boldeta)\|}{n\|h_i(\bbeta, \boldeta)\|} - 1\right) \| h_i(\bbeta, \boldeta)\|^2  \\ 
& \qquad \qquad - \frac{N(N/n-1)}{N^2} \left(\frac{1}{N}\sum_{i =1}^N \|h_i(\bbeta, \boldeta)\|^2\right) \\
&= \frac{1}{nN^2}\left(\sum_{i=1}^N \| h_i(\bbeta, \boldeta)\| \right)\left(\sum_{j=1}^N \|h_j(\bbeta, \boldeta)\|\right) - \frac{1}{N^2}\sum_{i=1}^N \| h_i(\bbeta, \boldeta)\|^2  \\ 
& \qquad \qquad - \frac{1}{nN}\sum_{i=1}^N \| h_i(\bbeta, \boldeta)\|^2 +  \frac{1}{N^2}\sum_{i=1}^N \| h_i(\bbeta, \boldeta)\|^2 \\
&= \frac{1}{n}\left[\left(\frac{1}{N} \sum_{i=1}^N \|h_i(\bbeta, \boldeta)\|\right)^2 - \left(\frac{1}{N}\sum_{i=1}^N \|h_i(\bbeta, \boldeta)\|^2\right)\right] \\ 
& \leq 0 \tag{Cauchy-Schwartz}
\end{align*}

Hence, $\text{Var}_\text{Pois}\left(\tilde{\bbeta}_\text{OSMAC} \big| (\mathbf{y}, \mathbf{X}) \right) \leq \text{Var}_\text{Strat}\left(\tilde{\bbeta}_\text{Neyman} \big| (\mathbf{y}, \mathbf{X}) \right)$ asymptotically in the case where stratification is uninformative. Note that this inequality is (asymptotically) strict whenever $\text{Var}(\|h_i(\bbeta, \boldeta)\|) \neq 0$, which is the case in any setting of interest.

\subsection{Further details on subsampling strategies 1-6}

In the simulation study and data example, we conducted subsampling of $n$ units according to the following strategies: 
\begin{itemize}
    \item Strategy 1: Case-control sampling using the true outcome $y$. Here, $n$ samples are collected with $\text{min}(n/2,\text{number of cases})$ cases selected via simple random sampling from the cases and then $n - \text{min}(n/2,\text{number of cases})$ controls are selected via simple random sampling from the controls, where cases and controls are defined by the true outcome $y$.
    \item Strategy 2: Case-control sampling using the surrogate outcome $s$. Same as Strategy 1, only cases and controls are defined by the surrogate outcome $s$.
    \item Strategy 3: Optimal individualized sampling using the OSMAC probabilities of \cite{wang2018optimal}, which assumes the true outcome $y$ is known. We implement this approach using Poisson sampling instead of their original with-replacement approach based on results from \cite{wang2021comparative}, which suggest that Poisson sampling is more efficient. We also independently verified this result (results not shown). Because Poisson sampling was used, the displayed sample sizes only represent the expected sample size for this strategy.  
    \item Strategy 4: Optimal individualized sampling with a pilot study of size $n_1$ using the two-step algorithm of \cite{marks2025optimal}. To implement this approach, we use the exact code provided by \cite{marks2025optimal}, which uses with-replacement sampling. 
    \item Strategy 5: Optimal Stratified sampling using Equation \ref{eq:NeymanAll}, assuming the true outcome $y$ is known for all units in the original sample and hence estimating $V_{h,k}$ with the variance of $\mathbf{H}(\hat{\bbeta}_\text{MLE}, \hat{\boldeta})$. This strategy is the stratified sampling analog to the \cite{wang2018optimal} approach in Strategy 3. Strata are formed by first taking the influence functions for $\beta_1, \beta_2,$ and $\beta_3$ from the full-data MLE and categorizing each observation as below the 0.2 quantile, above the 0.8, or between the two for each influence function. These quantile cut points are not likely to be perfectly optimal, but they are based on the heuristic that splitting a large group of normally distributed random variables at 0.2 and 0.8 quantiles leads to three smaller groups with approximately equal within-group variances. Final strata are then formed according to combinations of the binary outcome and these categories, leading to a maximum of $3^3*2 = 52$ strata. Then, Neyman allocation of influence functions is used to determine the allocation to strata. 
    \item Strategy 6: Optimal Stratified sampling with a pilot sample of size $n_1$ assuming $y$ is not initially known, using the pilot sample to estimate $V_{h,k}$, as described by \cite{Mcisaac2015} and \cite{Yang2025a}. This is an approximation of Strategy 5, where strata and the pilot wave allocation are defined by repeating Strategy 5 with the surrogate $s$ in place of $y$. Then, the remaining samples are selected using the influence functions computed using the true $y$ for the units in the pilot sample.
\end{itemize}

\subsection{Supplemental Figures}

\begin{figure}
\centerline{\includegraphics[width=6.25in]{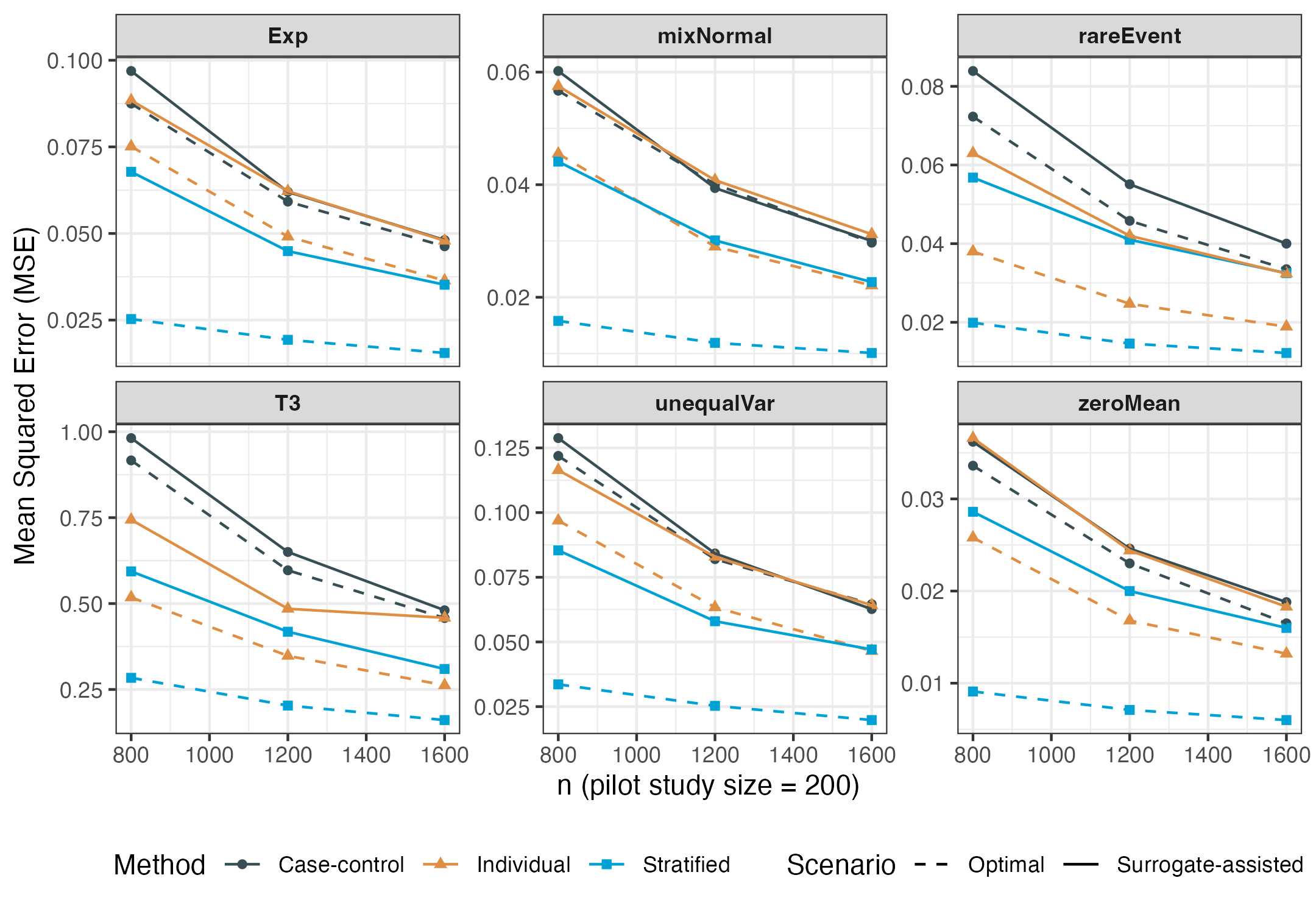}}
\caption{Empirical MSEs of each strategy under data-generating scenarios 1-6 with three covariates, a low level of misclassification for the surrogate, and a pilot study size of $n_1 = 200$. Dashed lines represent strategies that rely on $y$ being known, and solid lines represent strategies that use a surrogate and/or a pilot study. }
\label{fig:s1}
\end{figure}

\begin{figure}
\centerline{\includegraphics[width=6.25in]{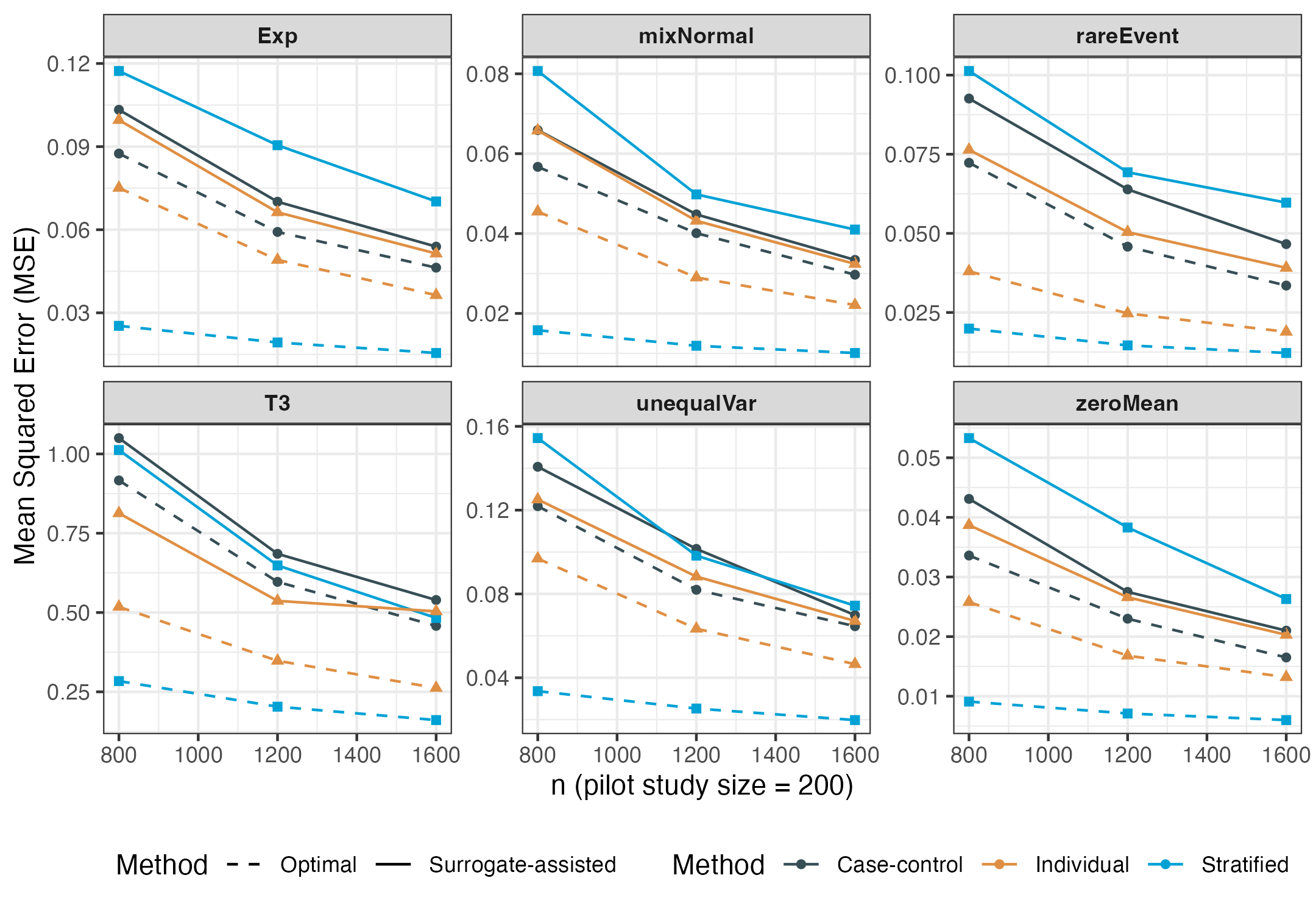}}
\caption{Empirical MSEs of each strategy under data-generating scenarios 1-6 with three covariates, a high level of misclassification for the surrogate, and pilot study size of $n_1$ = 200. Dashed lines represent strategies that rely on $y$ being known, while solid lines represent strategies that use a surrogate and/or a pilot study.}
\label{fig:s2}
\end{figure}

\begin{figure}
\centerline{\includegraphics[width=6.25in]{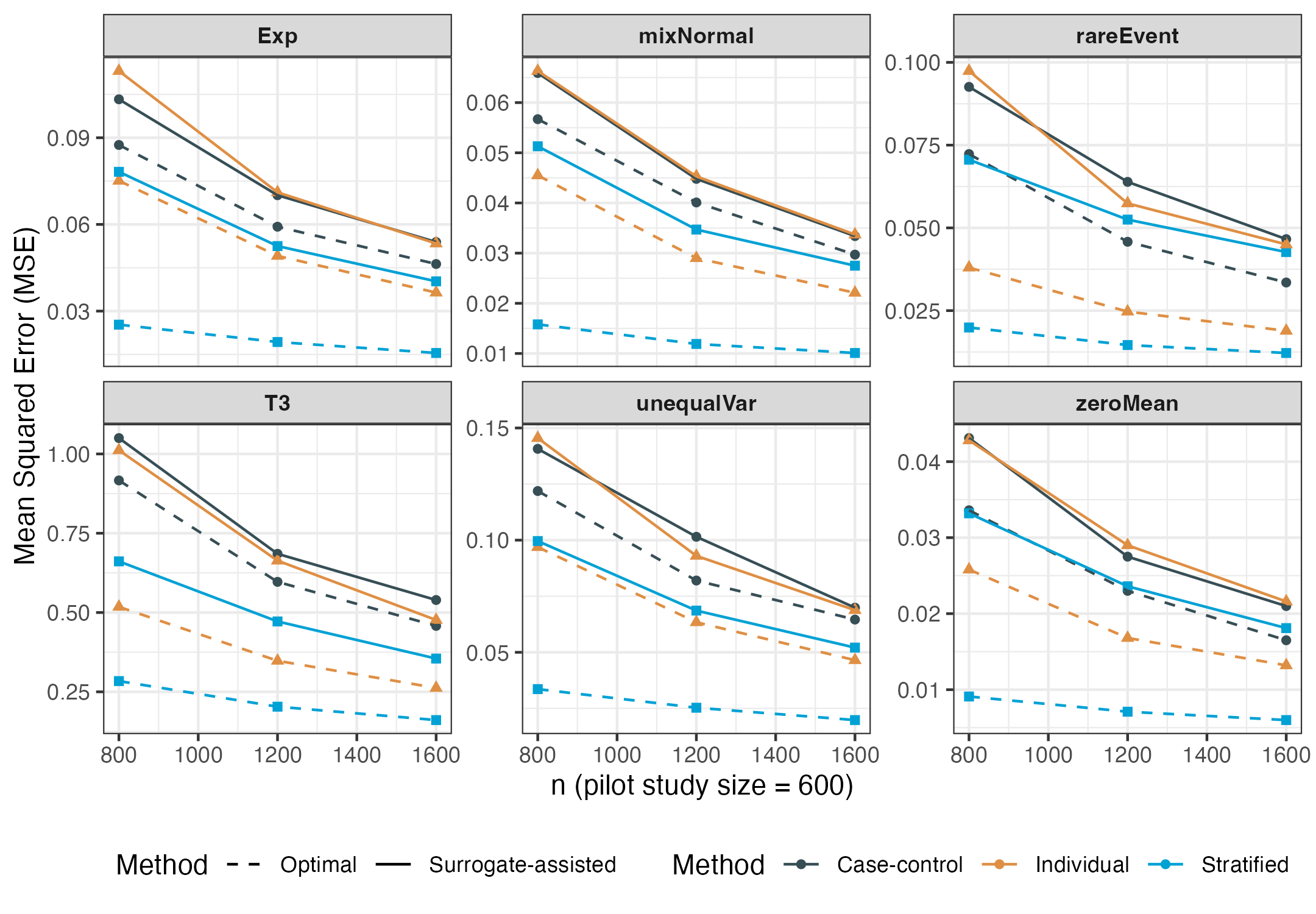}}
\caption{Empirical MSEs of each strategy under data-generating scenarios 1-6 with three covariates, a high level of misclassification for the surrogate, and a pilot study size of $n_1$ = 600. Dashed lines represent strategies that rely on $y$ being known, while solid lines represent strategies that use a surrogate and/or a pilot study. }
\label{fig:s3}
\end{figure}

\begin{figure}
\centerline{\includegraphics[width=6.25in]{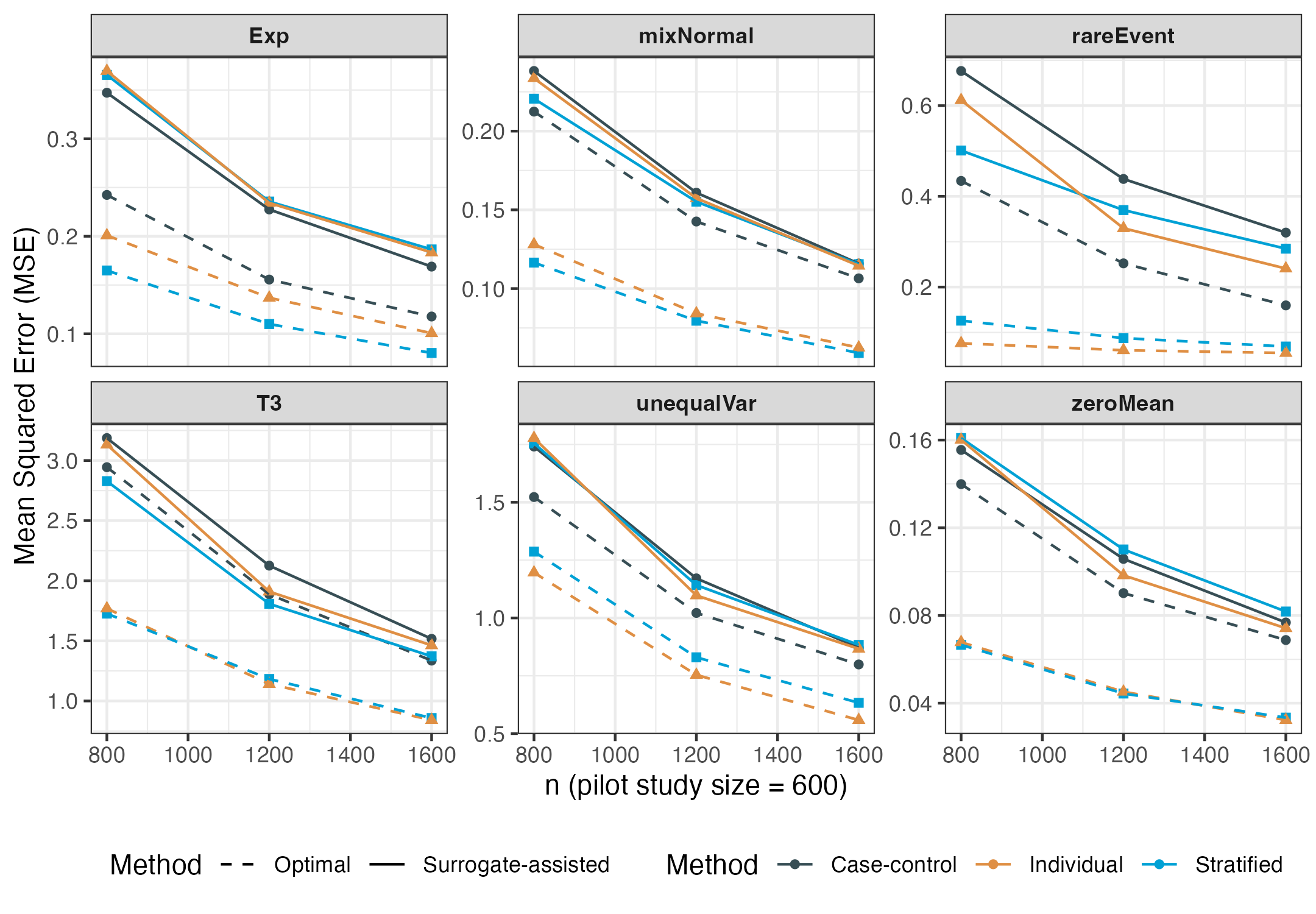}}
\caption{Empirical MSEs of each strategy under data-generating scenarios 1-6 with seven covariates, a high level of misclassification for the surrogate, and a pilot study size of $n_1$ = 600. Dashed lines represent strategies that rely on $y$ being known, while solid lines represent strategies that use a surrogate and/or a pilot study. }
\label{fig:s4}
\end{figure}

\renewcommand{\arraystretch}{1}

\newpage


\label{lastpage}
\end{document}